\pdfoutput=1
\documentclass[%
    aps,superscriptaddress,  
    prl, 
    reprint,
    a4paper,
    ]{revtex4-1} 

\usepackage{nicefrac}
\usepackage[english]{babel}

\usepackage[nosectionfonts,nonotetitle,nocosmeticdefs,nopagegeomdefs,%
    svn
    ]{docnote}
\SVN $Id: e6afed1da219a42e3790c4f4a79b47c74d83014e $
\SVN $Author: pfaist@ethz.ch $
\SVN $Date: 2015-02-16 19:08:39 +0100 (Mon, 16 Feb 2015) $

\renewcommand{\paragraph}[1]{\par\vspace{1ex}\emph{#1.---}\ignorespaces}
\def\subparagraph#1{\emph{#1} }

\makeatletter
\def\pdfstartlink@attr{attr{/Border[0 0 0 [1 5] ]/H/I/C[0 1 1]}}%
\def\@@Doi#1{\textcolor{docnotelinkcolor}{#1}\@@endlink}
\makeatother

\begin{document}

\title{Gibbs-Preserving Maps outperform Thermal Operations in the quantum regime}
\author{Philippe Faist}
\affiliation{Institute for Theoretical Physics, ETH Zurich, 8093 Switzerland}
\email[]{pfaist@phys.ethz.ch}
\author{Jonathan Oppenheim}
\affiliation{Department for Physics and Astronomy, University College of London, WC1E 6BT, U.K.}
\author{Renato Renner}
\affiliation{Institute for Theoretical Physics, ETH Zurich, 8093 Switzerland}
\date{\SVNDate}


\begin{abstract}
  In this brief note, we compare two frameworks for characterizing possible operations in
  quantum thermodynamics. One framework considers Thermal Operations---unitaries which
  conserve energy. The other framework considers all maps which preserve the Gibbs state
  at a given temperature.  Thermal Operations preserve the Gibbs state; hence a natural
  question which arises is whether the two frameworks are equivalent.  Classically, this
  is true---Gibbs-Preserving Maps are no more powerful than Thermal Operations. Here, we
  show that this no longer holds in the quantum regime: a Gibbs-Preserving Map can
  generate coherent superpositions of energy levels while Thermal Operations cannot. This
  gap has an impact on clarifying a mathematical framework for quantum thermodynamics.
\end{abstract}

\maketitle

The field of thermodynamics has recently seen a surge of activity~\cite{%
  Landauer1961_5392446Erasure,%
  Bennett1982IJTP_ThermodynOfComp,%
  Horodecki2003PRA_NoisyOps,%
  Janzing2000_cost,%
  Dahlsten2011NJP_inadequacy,%
  Popescu2006NPhys_entanglement,%
  Gemmer2009_quantum,%
  delRio2011Nature,%
  Brandao2013_resource,%
  Aberg2013_worklike,%
  Horodecki2013_ThermoMaj,%
  Egloff2012arXiv,%
  Faist2012arXiv,%
  Skrzypczyk2013arXiv_extracting,%
  Reeb2014NJP_improved,%
  Aberg2014PRL_catalytic,%
  Brandao2015PNAS_secondlaws,%
  NellyNg2014arXiv_limits,%
  Lostaglio2015NC_beyond,%
  Cwiklinski2014arXiv_limitations,%
  Renes2014EPJP_workcost%
}, in large part because of the application of techniques from information theory to the
subject. One of the key contributions has been a more precise definition of what
thermodynamics is, and this has allowed us to derive more rigorous quantitative statements
about the laws of thermodynamics. Traditionally, a number of processes such as isothermal
expansion or adiabatic processes were considered allowable thermodynamical operations, but
the precise nature of what was allowed was never defined. Thermodynamics was considered to
consist of crude control of systems, but as experimental control has improved, what
constitutes a thermal process and what is considered to be disallowed was unclear.

However, once we define the allowable processes that constitute the field of
thermodynamics, we can explore the implications. What's more, we can explore what happens
in regimes which had previously been difficult to study, in particular, we can gain a
better understanding of thermodynamics at the quantum level. In recent approaches to
thermodynamics, one defines what thermodynamics is by specifying a set of state
transformations which an experimenter is allowed to perform ``for free'', i.e. at no work
cost---such a framework is called a resource theory. However, there is more than one
possible way to formulate the resource theory and it is crucial that we understand which
ones are appropriate and under which circumstances. Among the various mathematical
frameworks proposed to model thermodynamical operations, two have proven particularly
useful, namely the resource theory of Thermal Operations and the Gibbs-Preserving Maps.
Classically, these two frameworks are equivalent. If a transition between initial and
final states block diagonal in their energy eigenbasis is possible by Gibbs-Preserving
Maps, then it is also possible via Thermal Operations~\cite{Horodecki2013_ThermoMaj}. One
might suppose that this equivalence holds for arbitrary quantum states. In this short
note, we show that this is not the case: Gibbs-Preserving Maps can perform transitions
which Thermal Operations are incapable of.

\paragraph{Thermal Operations}
The resource theory of Thermal Operations has been extensively exploited to understand
thermodynamics at the quantum level%
~\cite{Janzing2006Habil,
  Horodecki2003PRA_NoisyOps,Brandao2013_resource,Horodecki2013_ThermoMaj,
  Gour2013arXiv_resource}.
One is allowed to perform any arbitrary joint unitary
operation, on a system and a heat bath at a given temperature $T$, which conserves the
total energy on the joint state of the system and the bath.
Thermal Operations also include bringing in arbitrary systems which are in the Gibbs state
at temperature $T$ (with arbitrary Hamiltonians). Finally, Thermal Operations allow
subsystems to be discarded for free, regardless of their state. 
%
Observe that Thermal Operations cannot change the Gibbs state into any other
state~\cite{Janzing2006Habil,Brandao2013_resource,Horodecki2013_ThermoMaj}. What's more,
the Gibbs state is the only 
state which has this property
~\cite{Brandao2015PNAS_secondlaws}. Gibbs states are thus 
the only state which can be allowed for free---if any other state were allowed, arbitrary
state transformations would be possible.

Crucially, Thermal Operations are not capable of
generating coherent superpositions of energy levels: 
a Thermal Operation must, by definition, commute with the total Hamiltonian, and thus
cannot generate such a superposition starting from an energy eigenstate.

\paragraph{Gibbs-Preserving Maps}
In the framework of Gibbs-Preserving Maps, one allows to carry out any completely
positive, trace-preserving map on a system which preserves the Gibbs state at a given
temperature $T$ (or 
``Gibbs-Preserving Map'', for short). These maps are a natural quantum-mechanical
generalization of the stochastic matrices used to characterize the so-called
\emph{$d$-majorization} or \emph{mixing character}~\cite{Ruch1976_increasingmixing,
  Ruch1980JMAA,Joe1990JMAA,BookMarshall2010Inequalities,Renes2014EPJP_workcost}.
Technically, these operations are convenient to work with as being a Gibbs-preserving map
is a semidefinite constraint. Also, in any reasonable thermodynamical framework, a map
that does not preserve the Gibbs state must cost work; this fact makes Gibbs-Preserving
Maps a conservative choice of framework for proving fundamental limits.

Since a Thermal Operation preserves the Gibbs state, the state transformations
possible with Thermal Operations are necessarily included in those achievable with
Gibbs-Preserving Maps.  Is the converse true? It is in the classical case, i.e.\@ for
states which are block diagonal in their energy eigenbasis. 
This can be seen as follows. A necessary and sufficient condition for transitions via
Thermal Operations is thermo-majorization~\cite{Horodecki2013_ThermoMaj}, a partial order
which is a generalization of majorization~\cite{%
  HardyLittlewoodPolyaInequalities1952,%
  BookBhatiaMatrixAnalysis1997,%
  Uhlmann1971_Dichtematrizen,%
  BookMarshall2010Inequalities%
}. More precisely, transformations are completely characterized in terms of
thermo-majorization of the initial and final states' spectrum with respect to the Gibbs
state. Now given the existence of a Gibbs-Preserving Map, classic results about
majorization ensure that the initial state's eigenvalues thermo-majorize the final state's
ones, meaning there exists also a Thermal Operation performing the
transformation~\footnote{%
  On a side note, this does not imply that Gibbs-Preserving Maps are equivalent to Thermal
  Operations as channels even when acting on block diagonal states. Rather, they are only
  equivalent in terms of state transitions. In other words, while the same pair
  $(\textit{input state}, \textit{output state})$ can be achieved in both frameworks for
  block-diagonal states, the actual logical processes, %
  i.e.\ trace-preserving completely positive maps or channels, that one can perform,
  differ.  Note also that even for a given fixed input state the actual channel performed
  is in general relevant, and not only the input and output state, as the full information
  about the channel can be obtained by keeping a purification of the
  input~\cite{Faist2012arXiv}.
  Additionally, a classic example (for the trivial Hamiltonian $H=0$) of a map
  preserving the
  fully mixed state but which is not a Thermal Operation is the Choi-Jamiolkowski map of
  the two-party reduced state of
  the Aharonov~\cite[Note {[9]}]{Fitzi2001PRL_byzantine} or determinant state
  \unexpanded{$\ket{\mathcal{A}}_{ABC}=\frac1{\sqrt{6}}\left[\ket{012}+\ket{120}+\ket{201}-\ket{210}-\ket{102}-\ket{021}\right]$},
  which is up to a local unitary the same example as in~\cite{Landau1993_birkhoff}.
}\nocite{Fitzi2001PRL_byzantine,Landau1993_birkhoff}.

We now address the question of whether Gibbs-Preserving Maps are strictly more powerful
than Thermal Operations, on arbitrary, quantum, input states.
We show that this is the case, by exhibiting an
example of a Gibbs-Preserving Map that performs a transformation forbidden by Thermal
Operations.

\begin{figure}
  \centering
  \includegraphics[width=7cm]{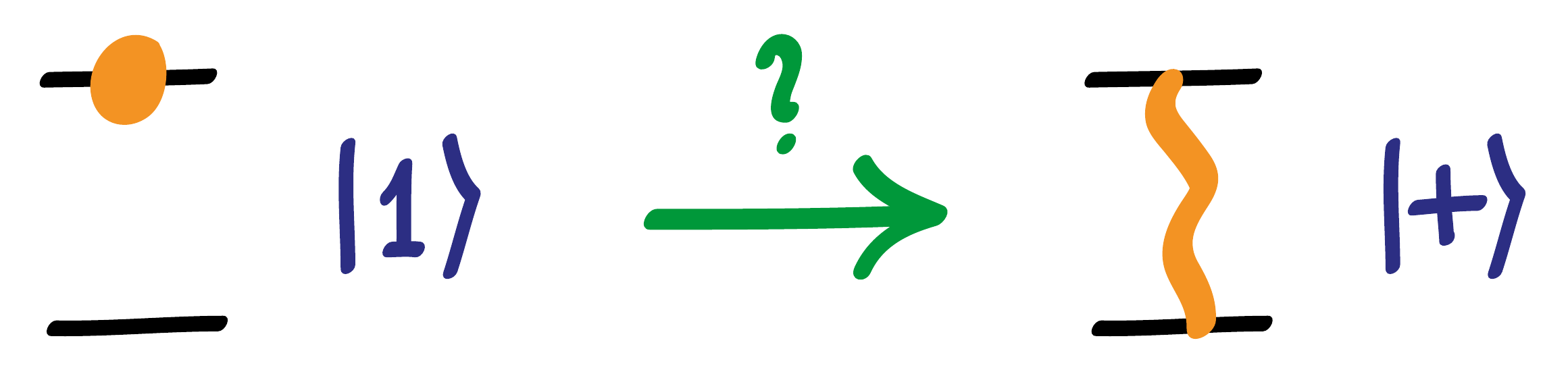}
  \caption{Problematic state transformation: If a qubit system is in a pure excited energy
    eigenstate $\ket1$, one would expect it is possible to bring it into any other state
    at no work cost, in particular in the coherent superposition of energy eigenstates
    $\ket+=\frac1{\sqrt2}[\ket0+\ket1]$. This is indeed possible with Gibbs-preserving
    Maps, however Thermal Operations forbid this transition because it requires nontrivial
    time control.}
  \label{fig:ProblematicTransition}
\end{figure}

\paragraph{The Example}
Consider a two-level system with an energy gap $\Delta E$. We denote the ground state by
$\ket0$ and the excited state by $\ket1$. Consider now the transformation:
\begin{align}
  \label{eq:thetransition}
  \ket 1\rightarrow\rho\ ,
\end{align}
where $\rho$ is any pure or mixed state. Depending on $\rho$, (in particular, in case
$\ket\rho=\ket{+}:=\frac12\left[\ket0+\ket1\right]$ as depicted in
Fig.~\ref{fig:ProblematicTransition}), this transformation needs to
``build'' coherence between the energy levels, which, as noted above, cannot be achieved
with Thermal Operations.
We now argue that, for any $\rho$, there exists nevertheless a Gibbs-Preserving Map
performing this transition.
Let $\beta$ be a fixed inverse temperature, and denote the Gibbs state on the system
by $\gamma=p_0\proj0+p_1\proj1$ with $p_0=1/Z$, $p_1 = \ee^{-\beta\Delta E}/Z$ and
$Z=1+\ee^{-\beta\Delta E}$. Let $\Phi$ be defined as
\begin{align}
  \label{eq:DefGibbsPresOpFromExcitedState}
  \Phi\left(\cdot\right) = \matrixel{0}{\cdot}{0}\,\sigma
  + \matrixel{1}{\cdot}{1}\,\rho\ ,
\end{align}
for some state $\sigma$ which we have not yet fixed. Note that $\Phi$ is completely
positive and trace-preserving. We also have $\Phi\left(\proj1\right)=\rho$ by
construction. The condition that $\Phi$ be Gibbs-preserving,
$\Phi\left(\gamma\right)=\gamma$, gives us
\begin{align*}
  p_0\sigma + p_1\rho = \gamma\ ,
\end{align*}
which implies
\begin{align}
  \label{eq:DefSigmaForGibbsPresOpFromExcitedState}
  \sigma = p_0^{-1}\left(\gamma - p_1\rho\right)\ .
\end{align}
This choice of $\sigma$ has unit trace, and is positive semidefinite; indeed, as
$\gamma\geqslant p_1\Ident$ (since $p_1$ is the smallest eigenvalue of $\gamma$) and
$\rho\leqslant\Ident$, we have $\gamma-p_1\rho\geqslant 0$.
This means that, with this choice of $\sigma$, $\Phi$ is precisely a completely positive,
trace-preserving, Gibbs-preserving channel which maps $\ket1$ to $\rho$. This map is forbidden
by Thermal Operations if $\rho$ contains
a  coherent superposition over energy levels, and we have the desired counter-example.

This example can easily be generalized to a system of $n$ arbitrary energy levels: if
$\ket n$, of energy $E_n$, is such that no other state has higher energy, a
Gibbs-Preserving Map $\Phi$ transforming $\ket n$ into any $\rho$ is given by
\begin{equation}
  \begin{aligned}
    \Phi\left(\cdot\right) &= \tr\left[\left(\Ident-\proj n\right)\,
      \left(\cdot\right)\right]\,\sigma
    + \tr\left[\proj n\,\left(\cdot\right)\right]\,\rho\ ,
  \end{aligned}
\end{equation}
where $\sigma = \left(\gamma-p_n\rho\right)/\left(1-p_n\right)$ and where the Gibbs state is
$\gamma=\sum p_i\proj i$ with $p_i=\ee^{-\beta E_i}/Z$ and $Z=\sum\ee^{-\beta E_i}$.

\paragraph{Discussion}
The observation of a gap between these two classes of operations leaves open the question
which of the two captures the actual physical situation. The Gibbs-Preserving Maps are
useful as the most permissive framework that is nontrivial; there is however no known
explicit microscopic model which corresponds to these operations. Furthermore, to observe
any coherence between energy levels one needs a time reference frame~\cite{%
  Bartlett2006IJQI_dialogue,%
  Molmer1997PRA_coherence,%
  Aharonov1967PR_charge,%
  Bartlett2007_refframes%
}, which might cost work to produce and eventually get degraded. Allowing the use of such a
resource catalytically enables operations that were otherwise forbidden~\cite{%
  Brandao2013_resource,%
  Aberg2014PRL_catalytic,%
  Brandao2015PNAS_secondlaws,%
  NellyNg2014arXiv_limits%
}, yet if the catalyst may be returned only approximately in its original state, then work
can be embezzled and all transformations are possible, rendering the framework
trivial~\cite{Aberg2014PRL_catalytic,Brandao2015PNAS_secondlaws,NellyNg2014arXiv_limits}.
Also, one usually expects
from a physical theory that one can ignore very unlikely events; this is by definition not
possible in the framework of exact catalysis. It is still an open question whether
transformations achievable by Gibbs-Preserving Maps coincide with Thermal Operations
combined with some form of time reference. On the other hand, if no such resource is
available, additional constraints related to time covariance are
required~\cite{Lostaglio2015NC_beyond}.

Our result, however, does not yet conclusively show that the Gibbs-Preserving Maps are
physically irrelevant. Intuitively, one could have argued from the start that the
transition~\eqref{eq:thetransition} should have been possible for any $\rho$: indeed, the
initial state has both maximal purity and highest possible energy. In fact, it is not
uncommon to assume that some form of coherence is available, for example in the context of
quantum computation, or, more generally, whenever a quantum system interacts with a
macroscopic system such as a detector or a laser field; the latter are usually modelled in
a coherent state.

Finally, it is worth noting that for Thermal Operations, there exists a set of conditions
which act as second laws, restricting which state transformations are
allowed~\cite{Brandao2015PNAS_secondlaws}. These take the form of a distance measure to the
Gibbs state, and are thus also a set of restrictions for Gibbs-Preserving Maps (due to the
data processing inequality for the R\'enyi relative entropies). 
As we now see that the two frameworks are inequivalent, this implies that a complete set
of second laws will necessarily involve functions which cannot be expressed in such a
form.

\paragraph{Acknowledgements}
The authors thank Johan \AA{}berg, Lea Kr\"amer, David Reeb, Joe Renes, L\'\i{}dia del
Rio, Paul Skrzypczyk, and Stephanie Wehner for helpful discussions.
PhF and RR are supported by the SNSF NCCR QSIT, the SNSF project 200020-135048, the ERC
grant 258932, and the COST Action MP1209. JO thanks the EPSRC for their support.
This work was in part developed during the ``Mathematical Horizons for
Quantum Physics~2'' program of IMS in Singapore and the ``Mathematical Challenges for
Quantum Information'' program of the Newton Institute in Cambridge, UK.

\bibliography{\jobname.bibolamazi}

\end{document}